\documentclass[conference]{IEEEtran}
\usepackage{cite}
\usepackage{amsmath,amssymb,amsfonts}
\usepackage{multirow}
\usepackage{subfigure}
\usepackage{booktabs}
\usepackage{algorithmic}
\usepackage{graphicx}
\usepackage{textcomp}
\usepackage{xcolor}
\usepackage{balance}
\usepackage{amsmath}
\def\BibTeX{{\rm B\kern-.05em{\sc i\kern-.025em b}\kern-.08em
    T\kern-.1667em\lower.7ex\hbox{E}\kern-.125emX}}
\begin{document}

\title{Learning to Compensate: A Deep Neural Network Framework for 5G Power Amplifier Compensation}

\author{\IEEEauthorblockN{Po-Yu Chen, Hao Chen, Yi-Min Tsai, Hsien-Kai Kuo, Hantao Huang, \\Hsin-Hung Chen, Sheng-Hong Yan, Wei-Lun Ou, Chia-Ming Cheng}
\IEEEauthorblockA{\textit{MediaTek Inc.} \\
Hsinchu, Taiwan}
}

\maketitle

\begin{abstract}
Owing to the complicated characteristics of 5G communication system, designing RF components through mathematical modeling becomes a challenging obstacle. Moreover, such mathematical models need numerous manual adjustments for various specification requirements. In this paper, we present a learning-based framework to model and compensate Power Amplifiers (PAs) in 5G communication. In the proposed framework, Deep Neural Networks (DNNs) are used to learn the characteristics of the PAs, while, correspondent Digital Pre-Distortions (DPDs) are also learned to compensate for the nonlinear and memory effects of PAs. On top of the framework, we further propose two frequency domain losses to guide the learning process to better optimize the target, compared to naive time domain Mean Square Error (MSE). The proposed framework serves as a drop-in replacement for the conventional approach. The proposed approach achieves an average of 56.7\% reduction of nonlinear and memory effects, which converts to an average of 16.3\% improvement over a carefully-designed mathematical model, and even reaches 34\% enhancement in severe distortion scenarios.
\end{abstract}


\section{Introduction}
In most communication systems, Power Amplifiers (PAs) are the major source of nonlinear and memory effects~\cite{JHK}, which cause severe spectral regrowth as shown in Figure 1. Spectral regrowth significantly degrades the signal quality due to high out-of-band power. Memory effects make the transmitted signal asymmetrically. These problems become critical issues of the next-generation 5G technologies. Moreover, in an edge device of 5G communication systems, both the high transmission power and the limited supply voltage make the non-linearity of PA even worse.

To overcome the aforementioned imperfections of PAs, Digital Pre-Distortion (DPD) has become a common and practical mechanism considering both integration complexity and effectiveness~\cite{Ghannouchi2009,wood2015,wood2017}. As illustrated in Figure 2, DPD generates additional signal distortions to compensate for the imperfections introduced by the PA. During the past decades, there has been an extensive study in none-learning based DPD, including Memory Polynomial (MP)~\cite{kim2001,Ding2004,Braithwaite2008}, Envelope-Memory Polynomial (EMP)~\cite{Hammi2008}, Generalized Memory Polynomial (GMP)~\cite{Morgan2006}, Dynamic Deviation Reduction (DDR) and Volterra Series (VS)~\cite{Schetzen2006}. Among these works, a mathematical PA model is firstly derived and the corresponding reverse function is then used to generate the pre-distortion signals.

Meanwhile, instead of deriving a mathematical model, learning-based approaches adopt neural networks to solve the problem. Learning-based approaches~\cite{Zhang2003,Devabhaktuni2001,JianjunXu2002,Mkadem2010,Liu2004,Ciminski2004,Mkadem2011,Gracia2019} demonstrate convincing results in which neural networks are trained to learn the pre-distortion signals for compensation. To the best of our knowledge, the existing learning-based compensations apply a straightforward training setup, in which Mean Square Error (MSE) loss or its variants are applied directly to the time domain signals. However, such time-domain MSE loss does not well characterize the severity of spectral regrowth in the frequency domain. Furthermore, in a real communication system, e.g. 3GPP's 5G specifications, there are explicit definitions and constraints on the characteristics of the frequency spectrum of transmitted signals~\cite{3GPP}. Overall, the time domain MSE loss might not only leads to sub-optimal results but also potentially breaks the communication specification.

To address the aforementioned shortages in the existing learning-based methods, we propose a learning-based framework for 5G PAs compensation. The framework differs from the existing works by introducing Deep Neural Networks (DNNs) to learn the 5G PAs' behaviors and characteristics. With the trained neural networks which represent the non-differentiable real PAs, the corresponding pre-distortion compensation is then learned in an end-to-end training paradigm.

In summary, our contributions are three-fold:
\begin{enumerate}
  \item We propose a end-to-end learning-based framework to jointly learn the behaviors, characteristics, and compensation of PAs.
  \item On top of the framework, we  propose two frequency domain losses to simultaneously minimize spectral regrowth and optimize toward communication specifications.
  \item According to the experiments on real 5G PAs, the proposed approach outperforms both the existing non-learning-based and learning-based methods by a significant margin. When compared to a widely used non-learning-based approach, our approach achieves 34\%, 11\%, and 4\% reductions of spectral regrowth at the corresponding supply voltages, $4.0V$, $4.2V$, and $4.6V$, respectively.
\end{enumerate}

The rest of this paper is organized as follows. First, the problem is formulated in Section~\ref{sec:problem_formulation}. The proposed framework and detailed working mechanism of PA modeling and DPD compensation are elaborated in Section~\ref{sec:proposed_framework}. The proposed frequency domain loss functions for model optimization are discussed in Section~\ref{sec:multi_objective_loss}. Section~\ref{sec:experiments} discusses the detail of dataset acquisition, experiment setup, comprehensive results, and analyses. Finally, the conclusion and future directions of this work are described in Section~\ref{sec:conclusions}.

\begin{figure}
\centering     
\subfigure[]{\label{fig:intro_a}\includegraphics[width=0.72\linewidth]{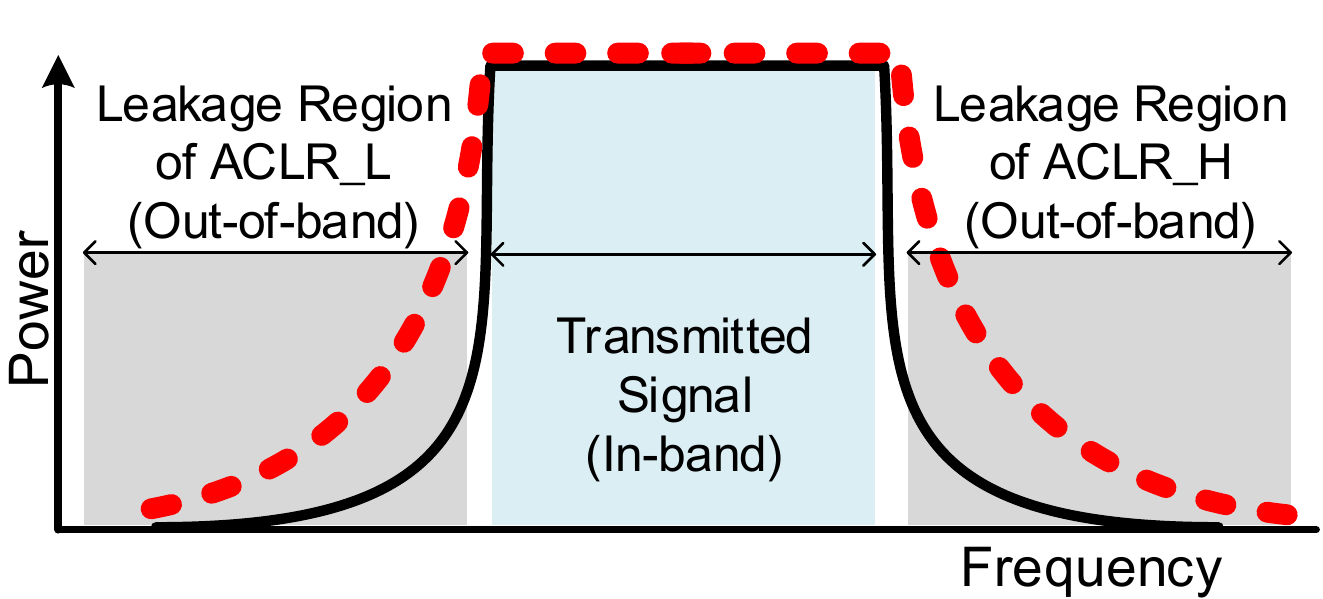}}
\subfigure[]{\label{fig:intro_b}\includegraphics[width=0.72\linewidth]{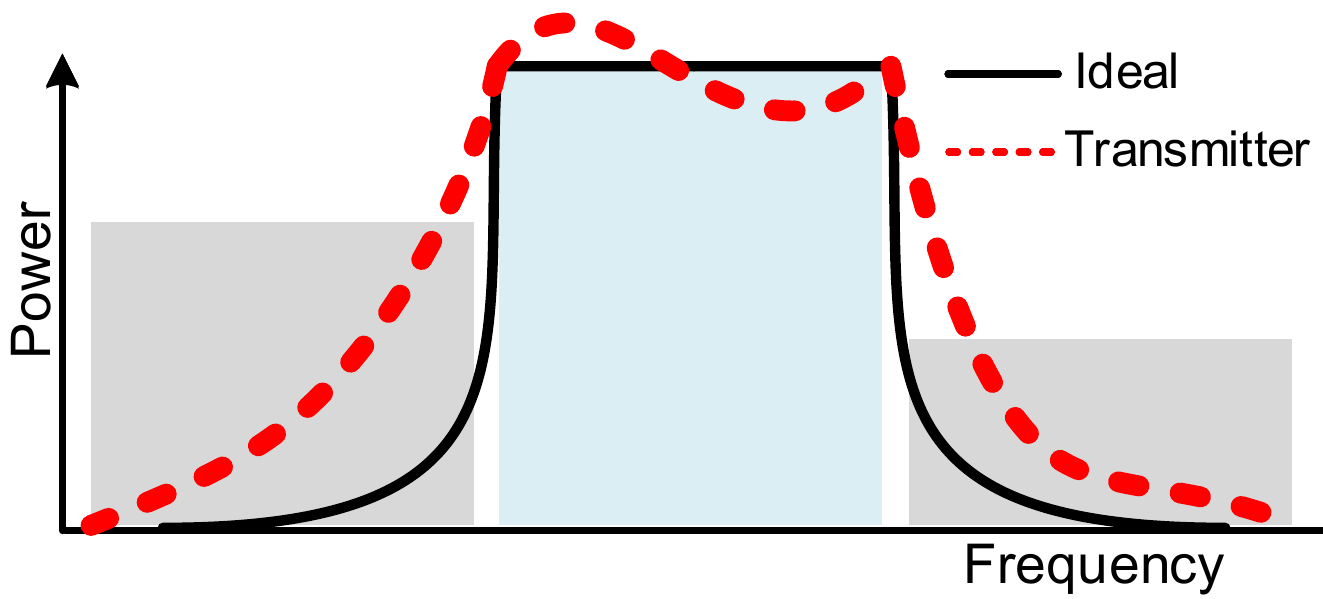}}
\caption{(a) Spectral regrowth due to non-linearity of PAs. (b) Memory effects cause in-band response and out-of-band spectral regrowth.}
\end{figure}

\section{Problem Formulation}
\label{sec:problem_formulation}

This paper addresses the DPD problem as represented in Figure 2. Consider a PA system defined by $y(n) = f_{PA}(x(n))$, where $x(n)$ and $y(n)$ indicate input and output signal, respectively, and $f_{PA}$ represents the PA function with certain nonlinearity. The problem can be formulated as

\begin{equation}
\label{eq::optimization}
    \min_{\theta} \ \ loss\big(y_{expect}(n), y_{DPD}(n)\big)
\end{equation}
where $y_{DPD}(n)$ is given by
\begin{equation}
\label{eq::dpd}
    y_{DPD}(n) = f_{PA}\big(f_{DPD}\big(x(n), \theta\big)\big)
\end{equation}

In Eq.~(\ref{eq::optimization}) and Eq.~(\ref{eq::dpd}), the objective is to find a compensation function $f_{DPD}$ to minimize the difference between the cascaded output $y_{DPD}(n)$ and an expected response $y_{expect}(n)$. Note that the difference is measured by using a distance-based loss function, e.g., L1 norm. According to Eq.~(\ref{eq::optimization}) and Eq.~(\ref{eq::dpd}), we present a learning-based framework in which DNNs are used to learn the compensator $f_{DPD}(\cdot, \theta)$ in which $\theta$ represents the weights for a neural network. The detail of the proposed learning-based framework will be introduced in Section~\ref{sec:proposed_framework}. On the other hand, loss function also plays an important role in guiding the learning process toward a better trajectory. The detail of the proposed loss functions will be elaborated in Section~\ref{sec:multi_objective_loss}.

\section{Proposed Framework}
\label{sec:proposed_framework}
In this section, we introduce the proposed pre-distortion compensation framework, which is trained end-to-end with the loss functions presented in Section~\ref{sec:multi_objective_loss}. The overall framework consists of two training stages. In the first stage, the \textit{Power Amplifier Network} (PAN) is trained to characterize nonlinear behaviors and memory effects of the non-differentiable real PAs (Figure~\ref{fig:pan_model}). Note that PA's behaviors heavily depend on the system parameters, such as voltage and frequency. Once the PAN training is finished, its weights are then fixed for being used in the next training stage. In the second training stage, the \textit{Pre-Distortion Network} (PDN) is end-to-end trained to obtain a compensation function with the differentiable PAN to mimic the behavior of a real PA (Figure~\ref{fig:pdn_model}). When all the training is finished, only the PDN is needed to be deployed in the system for inference.

\begin{figure}
\centering     
\subfigure[]{\label{fig:intro_c}\includegraphics[width=1.0\linewidth]{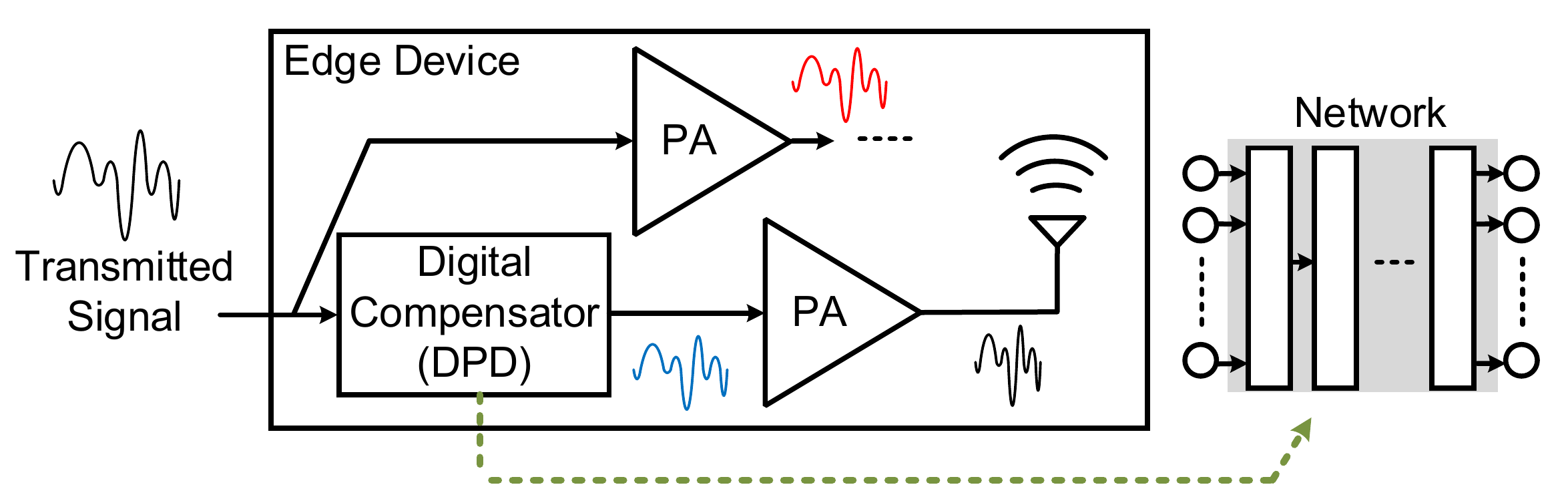}}
\subfigure[]{\label{fig:intro_d}\includegraphics[width=0.45\linewidth]{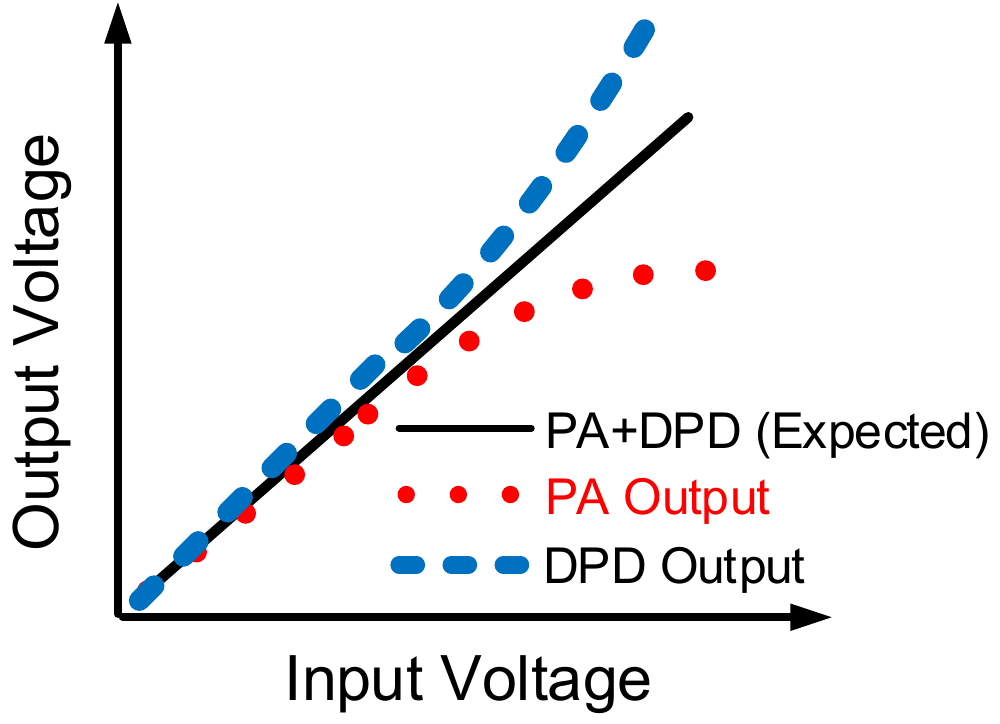}}
\caption{(a) Signal is distorted due to the nonlinear behaviors and memory effects of PAs (red signal). Pre-Distortion is applied to generate compensation signal (blue signal) to eliminate the nonlinear behaviors and memory effects of PAs (black signal). (b) An example of the compensation for nonlinear and memory effects of PAs .}
\end{figure}

\subsection{Power Amplifier Network}

The training data for PAN are obtained from multiple real PAs and processed as time sequence data with two channels which are noted as in-phase ($I$) and quadrature ($Q$) components (i.e., $x$ = \{$I_{t}$, $Q_{t}$\}, $t$ = 0, 1, …$T$ ). The generation of PA distortions under different conditions (i.e., voltage, frequency, or manufacturer) is formulated as learning a conditional distribution over the sequence data. We perform the conditioning by feeding \textit{Conditional PA Type} into PAN. The \textit{Conditional PA Type} is used as extra information and is concatenated along with the input sequence data $x$. With this training setting, a single PAN learns to generate any of the learned PA distortions under the given conditions. Similar conditional learning concepts can also be found in recent works, including Generative Adversarial Network (GAN) which learns disentangled salient information~\cite{Lample2017}, and a shared generator with conditional input vectors~\cite{Choi_2018_CVPR}.

\subsection{Pre-Distortion Network}

Given the trained PAN with fixed weights, PDN is then trained to learn the compensation signals for the distorted signals generated by the PAN. Instead of feeding the conditions into PDN, PDN is used to learn to extract the condition information directly from the signal pair of previous distorted and transmitted signals (dotted signals in Figure~\ref{fig:pdn_model}). In this way, PDN does not depend on the conditions (i.e., voltage, frequency, or manufacturer) and thus reduces the deployment complexity since no monitors are needed to obtain the system conditions. During training, we randomly select a condition input for PAN and train the PDN to generate the compensation signal. Be aware that amplitude normalization is applied to the ground truth and the compensated signal before the loss is calculated. To simplify the illustration, such amplitude normalization is omitted in Figure~\ref{fig:pdn_model}.

\section{Multi-Objective Loss Function}
\label{sec:multi_objective_loss}

In addition to the time domain MSE loss, this section introduces two proposed losses on the frequency domain. To address the complex requirements of 5G specification, this paper applies all the three loss functions to guide the optimization toward multiple objectives on both the time and frequency domains.

\subsection{Time Domain - MSE Loss ($tMSE$).} 
Fundamentally, measuring the difference between ground truth signal $x$ and predicted signal $\hat{x}$ is to compute the MSE (Eq. (\ref{eq::mse})) in time domain directly.  

\begin{equation}
\label{eq::mse}
    loss_{tMSE}(x,\hat{x})=\log \sum\| x - \hat{x} \|_{2}
\end{equation}

Minimizing the MSE can assess signal mismatch in the time domain. However, MSE is not sufficient to capture the mismatch in the frequency domain since a small signal distortion may induce tremendous deformation in the frequency domain. Therefore, it is critical to consider both time and frequency domain metrics. More details will be discussed in Section~\ref{sec:experiments}.

\begin{figure}
    \centering
    \includegraphics[width=0.46\textwidth]{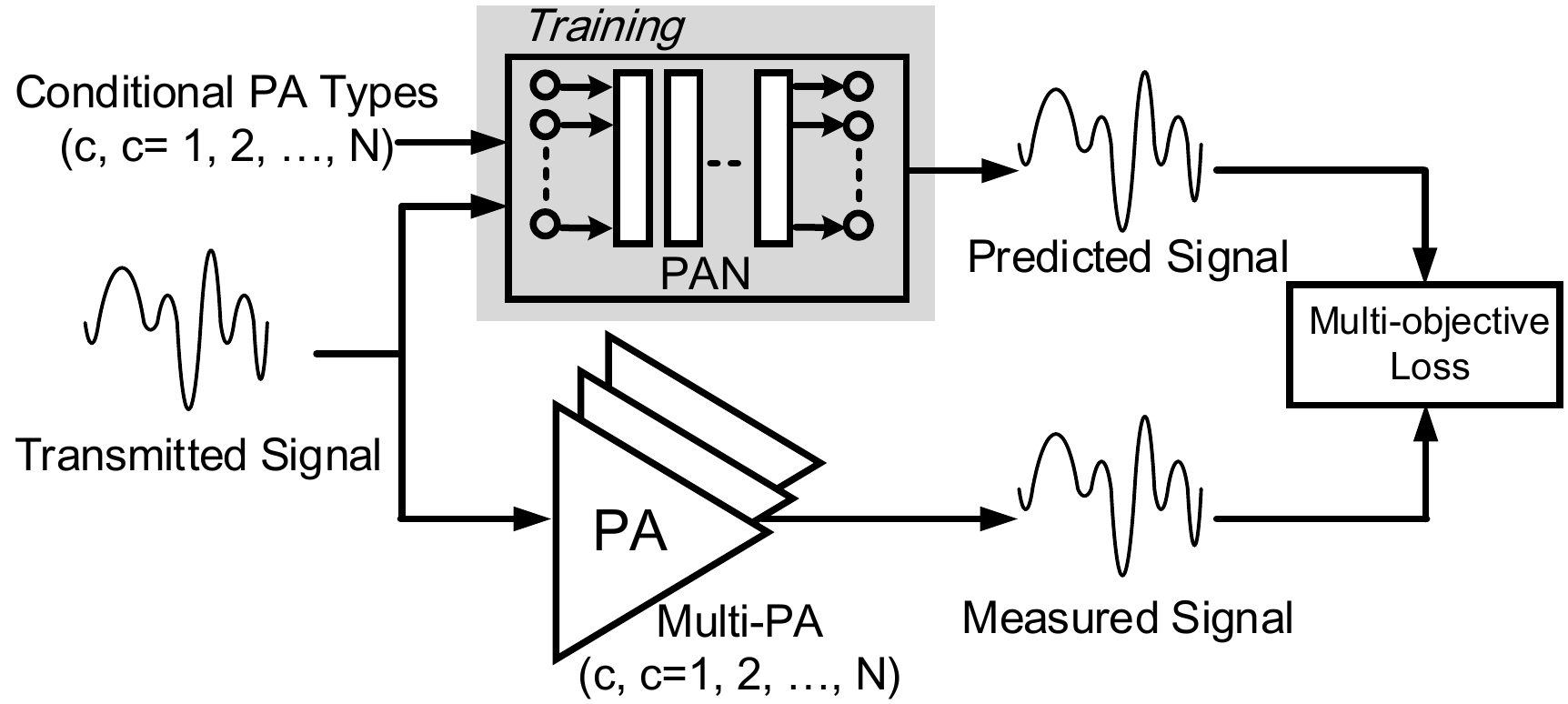}
    \caption{PAN training with conditional PA types.}
    \label{fig:pan_model}
\end{figure}

\subsection{Frequency Domain - MAE Loss ($fMAE$).} 
To obtain the frequency spectrum for this loss, we apply the Short Time Fourier Transform (STFT) on both $x$ and $\hat{x}$ and compute their absolute value in log scale. The detailed parameters are summarized in Section \ref{sec:experiments}. In the transformed spectrum, each complex number in an interval represents a specific frequency range. The absolute value describes power magnitude in the specific frequency point. Because the power magnitude of the transmitted signal is generally larger than the out-of-band signal, we take Mean Absolute Error (MAE) (Eq. (\ref{eq:sift})) instead of MSE to avoid over biasing on the in-band signals.

\begin{equation}
\label{eq:sift}
    loss_{fMAE}(x,\hat{x})=\log \sum\| STFT(x) - STFT(\hat{x}) \|_{1}
\end{equation}

\subsection{Frequency Domain - Specification Loss ($fSPEC$).}
In communication systems, it requires both in-band and out-of-band spectrum characteristics to fulfill certain specifications. In this paper, we consider 3GPP's  Adjacent Channel Leakage power Ratio (ACLR) requirement~\cite{3GPP} and propose a specification loss for this requirement. According to the 3GPP's specification, ``\textit{ACLR is the ratio of the filtered mean power centred on the assigned channel frequency to the filtered mean power centred on an adjacent channel frequency}"~\cite{3GPP}. The corresponding ACLR equation is as in Eq. (\ref{eq:aclr}). We then define a specification loss (Eq. (\ref{eq:aclrloss})) to minimize the ACLR difference between ground truth and predicted signals.

\begin{equation}
\label{eq:aclr}
ACLR(x) = \log(\frac{\sum_{in\_band}{} |STFT(x)|^2}{\sum_{out\_band}{} |STFT(x)|^2})
\end{equation}

\begin{equation}
\label{eq:aclrloss}
loss_{fSPEC}(x,\hat{x})=|ACLR(x) - ACLR(\hat{x})|
\end{equation}

\begin{figure}
    \centering
    \includegraphics[width=0.47\textwidth]{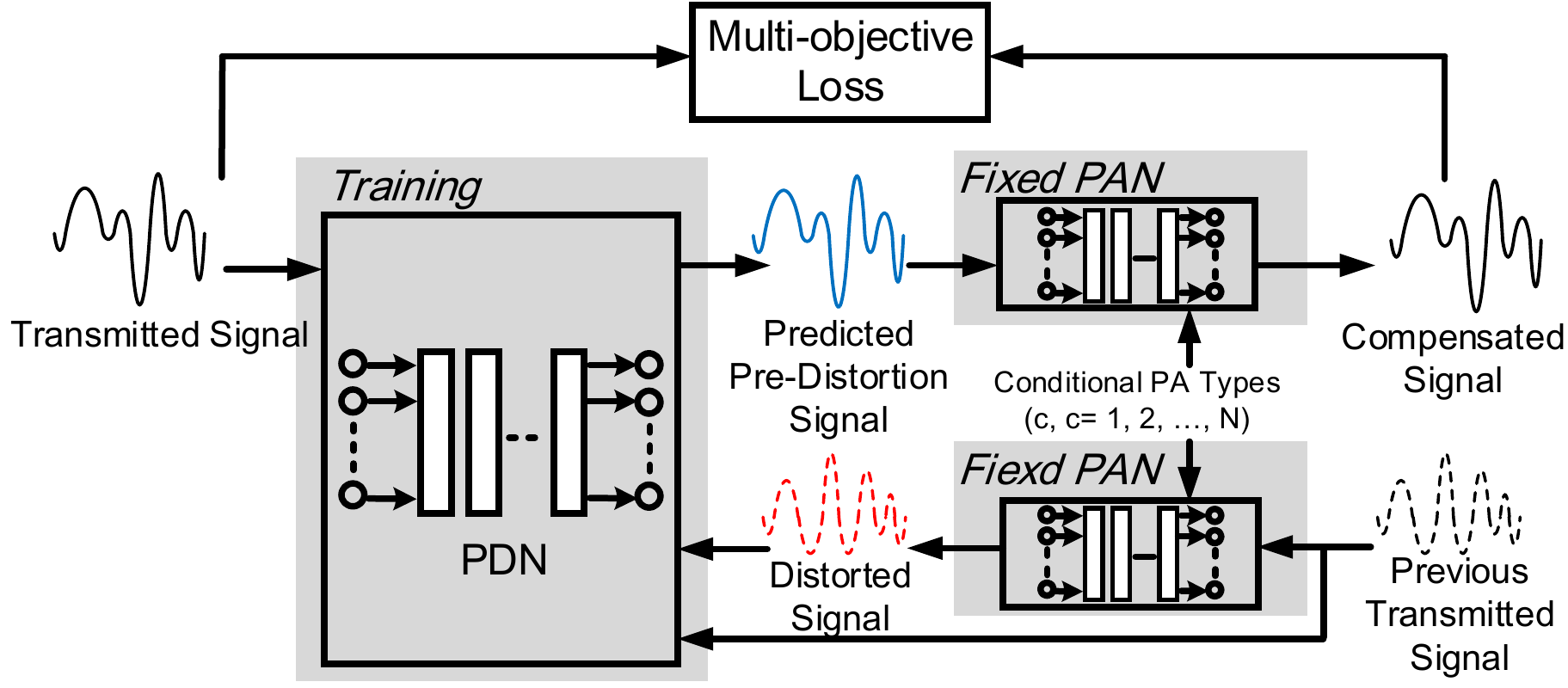}
    \caption{PDN training with pairs of previous signals (dotted).}
    \label{fig:pdn_model}
\end{figure}

\begin{figure*}
\centering     
\subfigure[]{\label{fig:pan_diff}\includegraphics[width=0.32\textwidth]{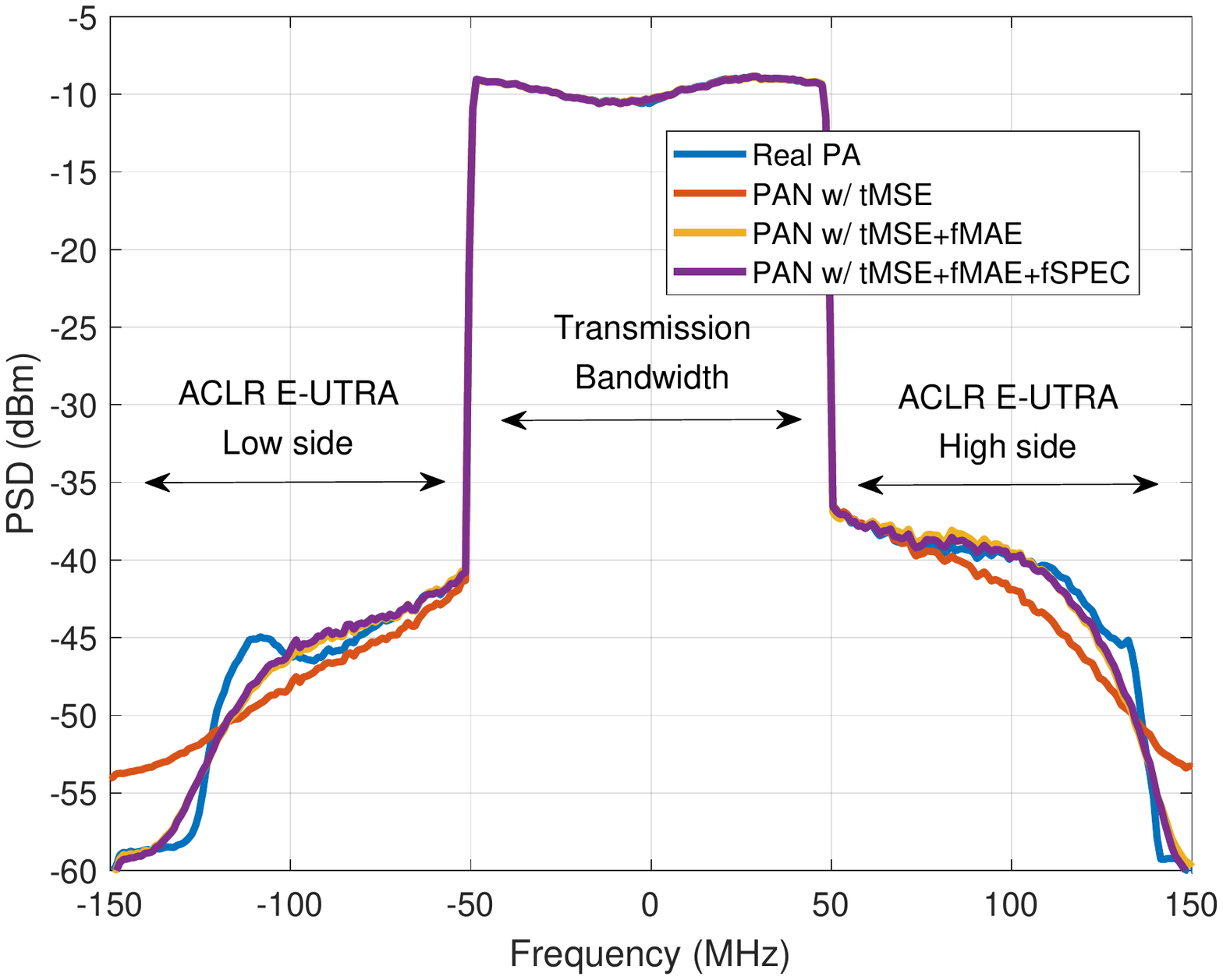}}
\subfigure[]{\label{fig:pdn_diff}\includegraphics[width=0.32\textwidth]{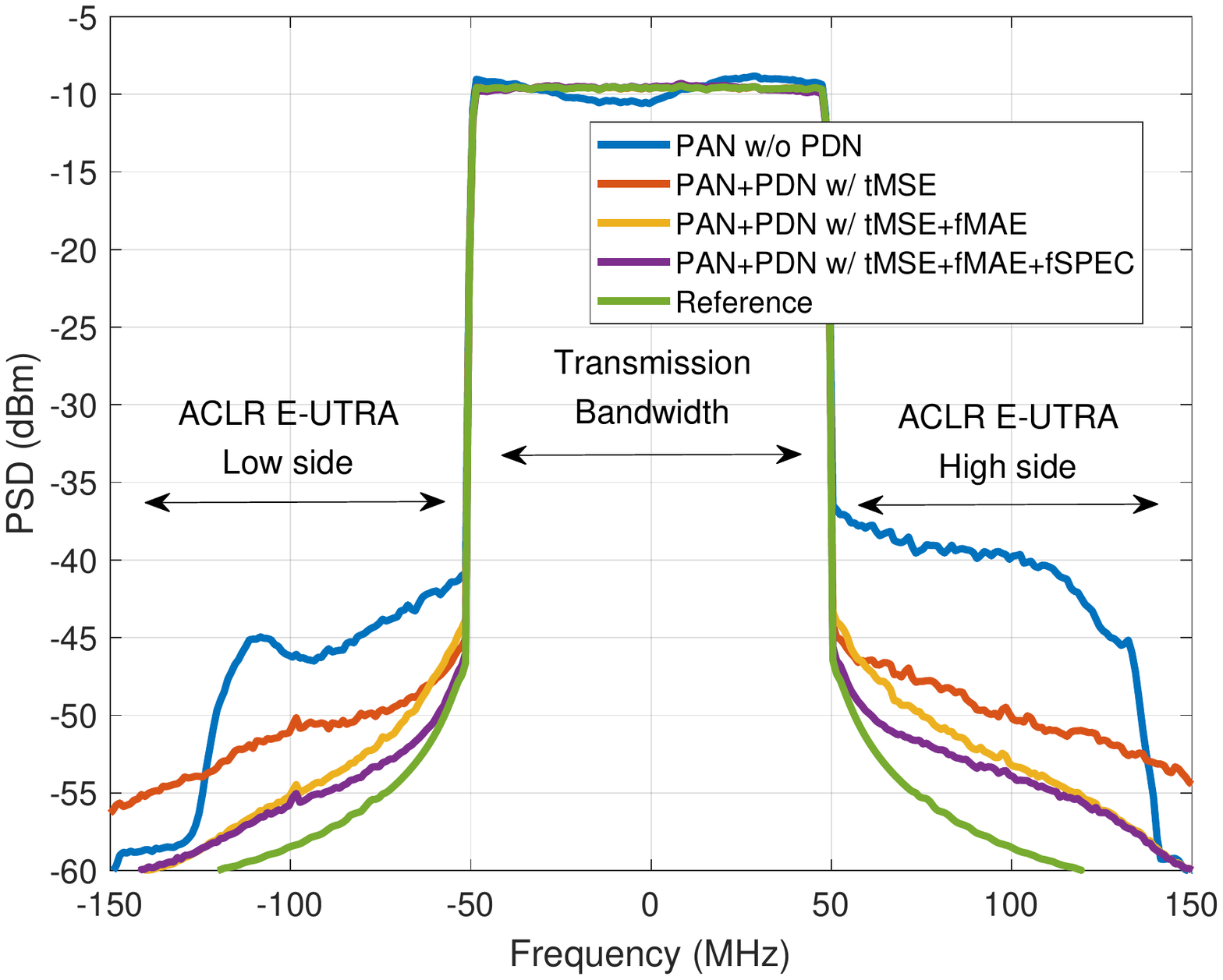}}
\subfigure[]{\label{fig:measure_output}\includegraphics[width=0.32\textwidth]{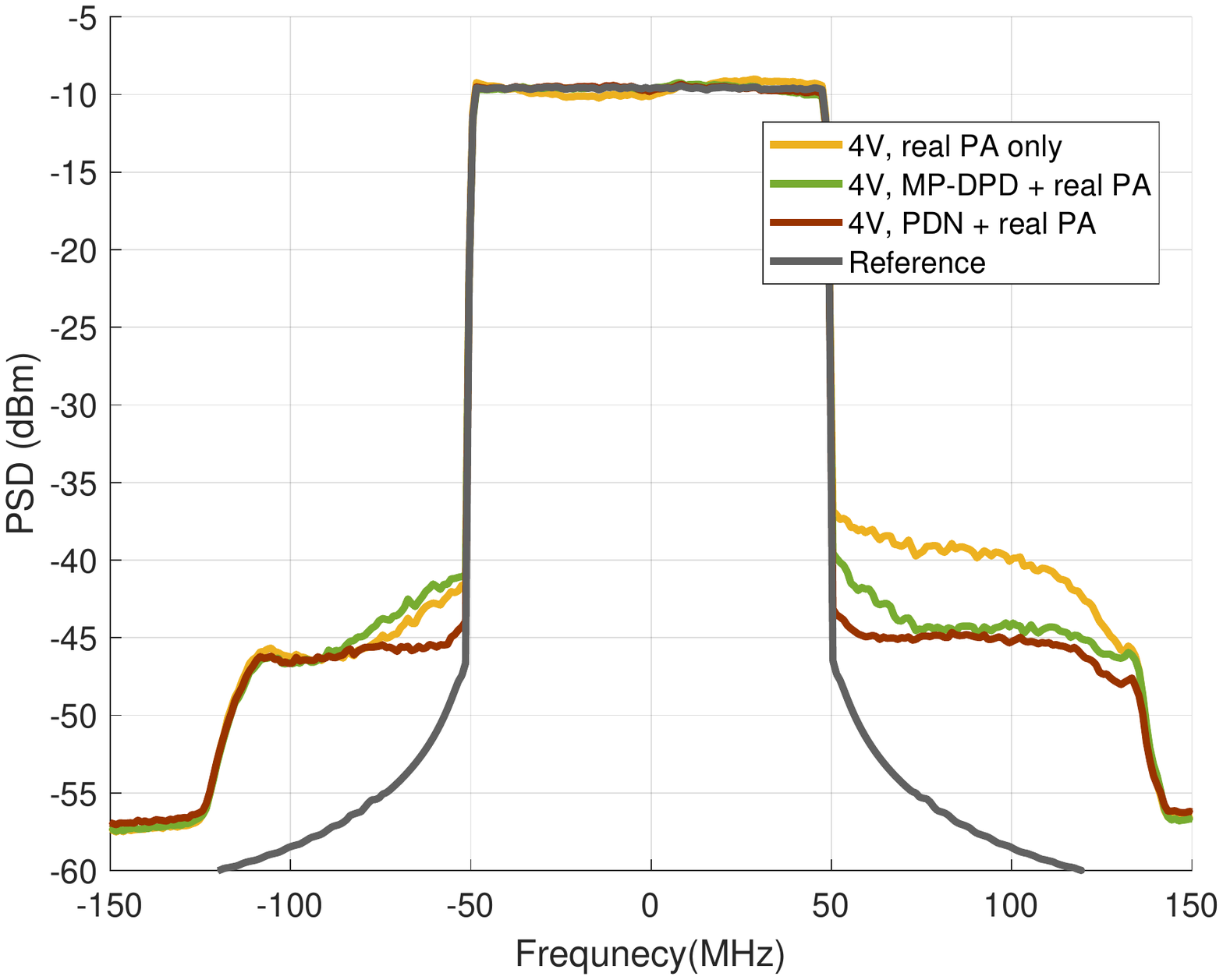}}
\caption{\label{fig:pan_diff_pdn_diff}
	    (a) Frequency spectrum of Real PA and PAN with different losses. (b) Frequency spectrum of PAN w/o PDN and PAN+PDN with different losses. (c) Frequency spectrum of real PA and different types of DPD methods.}
\end{figure*}

\begin{table*}[ht]
\caption{Performance of PAN which is trained to mimic the behavior and characteristics of Real PAs.}
\label{tab:pa_modeling}
\centering
\begin{tabular}{lccccc}
\toprule
\multicolumn{1}{c|}{} & \multicolumn{1}{c|}{Unit} & Real PA & PAN w/ tMSE & PAN w/ tMSE+fMAE & PAN w/ tMSE+fMAE+fSPEC\\
\midrule
\multicolumn{1}{l|}{MSE (time domain)} & \multicolumn{1}{c|}{dBc} &  N/A  & \textbf{-32.9} & -30.8 & -31.0\\
\multicolumn{1}{l|}{ACLR E-UTRA H}          & \multicolumn{1}{c|}{dB} & \textbf{30.9} & 31.8(+0.9) & 30.7(-0.2) & \textbf{30.9(+0)}\\
\multicolumn{1}{l|}{ACLR E-UTRA L}          & \multicolumn{1}{c|}{dB} & \textbf{36.7} & 37.8(+1.1) & \textbf{36.7(+0)} & \textbf{36.7(+0)}\\
\bottomrule
\end{tabular}
\end{table*}

\section{Experiments}
\label{sec:experiments}

This section details the experiment setups, including RF system setting and measurement setup for real PAs. The dataset, neural network structure, and training hyperparameters are also presented. We then discuss the experiment results of the proposed framework.

\subsection{System Details, Data Collection and Measurement}

This paper targets on a 5G communication system supporting New Radio (NR) with channel bandwidth 100MHz for mobile devices. In detail, the PAs operate at N41 band (2.496-2.690GHz) with sub-carrier spacing 30kHz and use QPSK modulation. Transmission Scheme is DFT-s-OFDM and transmission bandwidth is 98.28MHz and the high/low regions of leakage power (ACLR E-UTRA) are integrated over 98.28MHz centering at +/-100MHz offset respectively. We take 36 dB of ACLR E-UTRA which is higher than the requirement in 3GPP specification ~\cite{3GPP}.

The measurement and collection bench is built on NI PXIe-5840, consisted of a Vector Signal Generator (VSG) and a Vector Signal Analyzer (VSA). The VSG generates the modulated signal with up-conversion to RF and then feeds to the real PA for sample collection. The VSA down-converts RF signal to analog baseband and records the data. Signal timing and power normalization are further processed by MATLAB. To validate the PDN results, the output of PDN is firstly transferred from digital to analog signal by VSG and then fed to real PAs.

With the collection bench, we use Pseudo Random Number Generator to generate the sequence for both training and testing. We collect a complete transmission frame for each sequence, around 1.8M $I$ (In-phase) and $Q$ (Quadrature) with 368.68 MHz/sec sampling rate. The training data are collected with supply voltages, $4.0V$, $4.2V$ and $4.6V$, and frequency at 2.593GHz with a transmission power of 24 dBm. For testing data, samples with both 2.593GHz and 2.643GHz frequencies are further collected.

\subsection{Neural Network Structure and Training}

We take an identical network structure for both PAN and PDN. The network is composed of 6 convolution layers with an input sequence of 128 sample points (input length). In each sample, two input channels are used for $I/Q$. Each layer is a 1-D convolution with a kernel size of 3 and a stride of 1, and the output channel number for each layer is 16-32-64-64-128-2. Except for the last output layer, each convolution layer is followed by a batch normalization layer and PReLU activation function. Both PAN and PDN are trained on one Nvidia GTX980Ti. Adam optimizer is used with batch size 512 and 256 for PAN and PDN respectively. All models are trained for 4000 epochs with an initial learning rate of 0.01, and divided by 10 at every 1000 epochs.

\subsection{Results}

In this section, the proposed method is compared to two baselines. One is a learning-based baseline (implemented as the proposed method with only tMSE loss) \cite{Zhang2003,Devabhaktuni2001,JianjunXu2002,Mkadem2010,Liu2004,Ciminski2004,Mkadem2011,Gracia2019} and the other one is Memory Polynomial based baseline (MP-DPD, non-learning-based) \cite{kim2001,Ding2004,Braithwaite2008}. At the same time, ablation studies are also presented to support our contributions.

\begin{table*}[ht]
\small
\caption{Performance of PDN which generates the compensation signal for trained PAN.}
\label{tab:pd_modeling}
\centering
\begin{tabular}{lccccc}
\toprule
\multicolumn{1}{c|}{} & \multicolumn{1}{c|}{Unit} & w/o PDN & PDN w/ tMSE & PDN w/ tMSE+fMAE & PDN w/ tMSE+fMAE+fSPEC\\
\midrule
\multicolumn{1}{l|}{MSE (time domain)} & \multicolumn{1}{c|}{dBc} &  N/A  & \textbf{-35.5} & -28.9 & -28.3\\
\multicolumn{1}{l|}{ACLR E-UTRA H}          & \multicolumn{1}{c|}{dB} & 30.9 & 40.7 (+9.8) & 43.1 (+12.2) & \textbf{46.4 (+15.5)}\\
\multicolumn{1}{l|}{ACLR E-UTRA L}          & \multicolumn{1}{c|}{dB} & 36.7 & 42.8 (+6.1) & 45.5 (+8.8) & \textbf{49.6 (+12.9)}\\
\multicolumn{1}{l|}{\begin{tabular}{@{}l@{}}Out-of-band Spectrum Reduction\end{tabular}} & \multicolumn{1}{c|}{\%} &  0  & 89.4 & 93.9 & \textbf{97.2}\\
\bottomrule
\end{tabular}
\end{table*}

\begin{table}[t]
\small
\caption{Performance at different supply voltages of real PAs.}
\label{tab:diff_vol}
\centering
\setlength{\tabcolsep}{3pt}
\begin{tabular}{lrrrrrr}
\toprule
\multicolumn{1}{c|}{} & \multicolumn{2}{c|}{4.0 V} & \multicolumn{2}{c|}{4.2 V} & \multicolumn{2}{c}{4.6 V} \\
\midrule
\multicolumn{1}{l}{}  & \multicolumn{6}{c}{MP-DPD + Real PA} \\
\midrule
\multicolumn{1}{l|}{with DPD}         & No & \multicolumn{1}{r|}{Yes} & No & \multicolumn{1}{r|}{Yes} & No & Yes \\
\multicolumn{1}{l|}{MSE (dBc)}        & -20.4 & \multicolumn{1}{r|}{\textbf{-30.4}} & -21.4 & \multicolumn{1}{r|}{-25.9} & -23.0 & \textbf{-31.8} \\
\multicolumn{1}{l|}{ACLR E-UTRA H (dB)}   & 30.9 & \multicolumn{1}{r|}{33.1} & 34.3 & \multicolumn{1}{r|}{*36.3} & 36.1 &  *38.0\\
\multicolumn{1}{l|}{ACLR E-UTRA L (dB)}   & 36.7 & \multicolumn{1}{r|}{34.9} & 38.6 & \multicolumn{1}{r|}{*38.7} & 38.8 & *39.2 \\
\multicolumn{1}{l|}{\begin{tabular}{@{}l@{}}Out-of-band \\ Spectrum Reduction (\%)\end{tabular}} & & \multicolumn{1}{r|}{39} & & \multicolumn{1}{r|}{42} & & 40 \\
\midrule
\multicolumn{1}{l}{}  & \multicolumn{6}{c}{PDN + Real PA} \\
\midrule
\multicolumn{1}{l|}{with DPD}        & No & \multicolumn{1}{r|}{Yes} & No & \multicolumn{1}{r|}{Yes} & No & Yes \\
\multicolumn{1}{l|}{MSE (dBc)}        & -20.4 & \multicolumn{1}{r|}{-29.5} & -21.4 & \multicolumn{1}{r|}{\textbf{-30.3}} & -23.0 & -31.6 \\
\multicolumn{1}{l|}{ACLR E-UTRA H (dB)}   & 30.9 & \multicolumn{1}{r|}{\textbf{*36.7}} & 34.3 & \multicolumn{1}{r|}{\textbf{*37.4}} & 36.1 & \textbf{*38.3} \\
\multicolumn{1}{l|}{ACLR E-UTRA L(dB)}   & 36.7 & \multicolumn{1}{r|}{\textbf{*38.7}} & 38.6 & \multicolumn{1}{r|}{\textbf{*39.5}} & 38.8 & \textbf{*39.9} \\
\multicolumn{1}{l|}{\begin{tabular}{@{}l@{}}Out-of-band \\ Spectrum Reduction (\%)\end{tabular}} & & \multicolumn{1}{r|}{\textbf{73}} & & \multicolumn{1}{r|}{\textbf{53}} & & \textbf{44} \\
\bottomrule
\multicolumn{7}{r}{*meet industrial ACLR requirement (36 dB)} \\
\end{tabular}
\end{table}

\subsubsection{Learning PAs with Multi-Objective Optimization} 

Figure~\ref{fig:pan_diff} shows the resultant frequency spectrum of the three loss configurations for PAN. As depicted, in the region of transmission bandwidth, all configurations perform similarly well to characterize the real PA's spectrum. However, in the regions of ACLR E-UTRA (both Low and High), since real PA's non-linearity causes severe spectral regrowth, both the two proposed losses ($fMAE$ and $fSPEC$) carry out closer characteristics compared to real PA. As shown in Table~\ref{tab:pa_modeling}, although applying the time domain loss $tMSE$ achieves the lowest MSE error, it does not well represent the corresponding ACLR E-UTRA characteristics for real PAs (30.9 dB and 36.7 dB). On the other hand, applying the frequency domain losses, $fMAE$, and $fSPEC$, has a better match of the ACLR E-UTRA characteristics for real PAs. When combining all the three losses, the result perfectly matches the real PA's characteristics (30.9 dB and 36.7 dB) with only a slight increment in MSE.

\subsubsection{Learning DPD Functionality}
In Figure \ref{fig:pdn_diff}, PDN is applied to compensate for power spectrum reduced by the PA. As shown in the figure, the frequency domain losses, $fMAE$ and $fSPEC$, lead to a larger error in MSE. However, such frequency domain and communication specification consideration make the PDN better perform on the compensation functionality. As listed in Table~\ref{tab:pd_modeling}, when combining all the three losses, the resultant PDN network achieves 97\% out-of-band spectrum reduction compared to the one without DPD functionality. Note that such compensated signal is very close to the perfect reference signals as illustrated in Figure \ref{fig:pdn_diff}.

\subsubsection{Evaluation on Real PAs}

Although PDN can provide impressive performance on a system with PAN (mimics real PAs), the mismatch between PANs and real PAs may degrade the performance of compensation. Therefore, we further feed the pre-distorted signal from PDN into real PAs to estimate the compensation performance. As shown in Figure~\ref{fig:measure_output}, the proposed PDN effectively reduces spectral regrowth in both in-band and out-of-band regions. As listed in Table~\ref{tab:diff_vol}, a conventional and widely adopted DPD method, Memory Polynomial (MP-DPD), is implemented as a baseline~\cite{Morgan2006}. The memory depth and polynomial order used for comparison are respectively 3 and 5 in MP-DPD. According to Table~\ref{tab:diff_vol}, PDN achieves 73\%, 53\%, and 44\% reduction of out-of-band spectrum regrowth at supply voltage $4.0V$, $4.2V$, and $4.6V$, respectively. The proposed PDN steadily outperforms the MP-DPD baseline, especially in the lower supply voltage which introduces severe nonlinear problems and memory effects.

\begin{table}[t]
\small
\caption{Performance of Unseen Cases.}
\label{tab:unseen}
\centering
\begin{tabular}{lcrrrr}
\toprule
\multicolumn{1}{c|}{} & \multicolumn{1}{c|}{Unit} & \multicolumn{2}{c|}{Unseen 1} & \multicolumn{2}{c}{Unseen 2} \\
\midrule
\multicolumn{1}{l|}{with DPD}        & \multicolumn{1}{c|}{} & No & \multicolumn{1}{r|}{Yes} & No & Yes \\
\multicolumn{1}{l|}{Operating Frequency}  & \multicolumn{1}{c|}{GHz} & \multicolumn{2}{c|}{2.593} & \multicolumn{2}{c}{2.643} \\
\multicolumn{1}{l|}{Transmission Power}      & \multicolumn{1}{c|}{dBm} & \multicolumn{2}{c|}{25} & \multicolumn{2}{c}{24} \\
\multicolumn{1}{l|}{MSE}        & \multicolumn{1}{c|}{dBc} & -19.7 & \multicolumn{1}{r|}{-28.1} & -23.1 & -21.3 \\
\multicolumn{1}{l|}{ACLR E-UTRA H}   & \multicolumn{1}{c|}{dB} & 30.9 & \multicolumn{1}{r|}{33.7} & 38.9 & 43.1 \\
\multicolumn{1}{l|}{ACLR E-UTRA L}   & \multicolumn{1}{c|}{dB} & 37.3 & \multicolumn{1}{r|}{38.1} & 37.2 & 37.6 \\
\multicolumn{1}{l|}{\begin{tabular}{@{}l@{}}Out-of-band \\ Spectrum Reduction\end{tabular}} & \multicolumn{1}{c|}{\%} & &
\multicolumn{1}{r|}{\textbf{48}} & & \textbf{9} \\
\bottomrule
\end{tabular}
\end{table}

\subsubsection{Extend to Unseen Cases}
We further evaluate the PDN on two unseen data ($Unseen 1$ and $Unseen 2$) as shown in Table~\ref{tab:unseen}. In $Unseen 1$, the PA operates at 2.593GHz with larger transmission power, 25 dBm, which introduces even severe nonlinear problems and memory effects. In $Unseen 2$, the PA operates at 2.643GHz with similar transmission power, 24 dBm, which introduces a different PA behavior and characteristic. In both unseen cases, PDN still effectively achieves positive reductions of out-of-band spectrum regrowth as shown in Table~\ref{tab:unseen}.

\subsubsection{Discussion on Network Architecture}

Comparing to the existing learning-based approaches, most of the works adopt parameter-intensive neural network architectures, such as Multi-Layer Perceptron (MLP)~\cite{Zhang2003,Devabhaktuni2001,JianjunXu2002,Mkadem2010,wood2015,wood2017}, Recurrent Neural Networks (RNN)~\cite{Ciminski2004}, and Time-Delay Neural Networks (TDNN)~\cite{Liu2004}. Comparing to these architectures, Convolutional Neural Network (CNN) is much computation and parameter efficient for being embedded in practical systems. CNNs have also been widely studied in computer vision tasks, including classification, detection, segmentation, etc. Recently, CNNs have also demonstrated the capability to cope with sequence analyses. For example, the long-term dependency of a sequence is captured by stacking multiple convolutions with large receptive fields in WaveNet~\cite{wavenet2016,Oord2018ParallelWF}.

\section{Conclusions and Future Work}
\label{sec:conclusions}
This paper presents a framework to learn the characteristics of 5G PAs with different characteristics (voltage, frequency, and transmission power). Moreover, correspondent DPD signals are also learned to compensate for the nonlinear and memory effects of 5G PAs. On top of the framework, we further propose two frequency domain losses, $fMAE$ and $fSPEC$, to minimize spectral regrowth and optimize toward communication specifications at the same time. According to the experiments on real 5G PAs, the proposed approach outperforms both the non-learning- and learning-based existing works by a large margin. Without losing the generality, the two proposed frequency domain losses ($fMAE$ and $fSPEC$) can also be applied to other applications, especially in the domain of high-frequency complex signals. In these domains, it is common to consider (and/or measure if applicable) both time and frequency domain co-optimization. Such applications include style transfer in text-to-speech~\cite{jia2018} and voice cloning~\cite{Arik2018} in which the frequency characteristic is very important (to capture the style of voice). In addition to the speech/audio domain, the same idea can also be applied to medical signal processing (e.g. ECG/EKG, EEG, EMG signals). In addition, frequency domain losses can also be used in communication systems, such as Wi-Fi channel state information (CSI), which has applications like indoor localization~\cite{Wang2017} and human activity recognition~\cite{Adib2015}.

\balance
\bibliographystyle{IEEEbib}
\bibliography{refs}

\end{document}